# Millimeter Wave LOS Coverage Enhancements with Coordinated High-Rise Access Points

Yinan Qi, Mythri Hunukumbure and Yue Wang
Samsung Electronics R&D Institute UK, Staines, Middlesex TW18 4QE, UK
{yinan.qi, mythri.h, yue2.wang}@samsung.com

*Abstract*— Millimetre wave (mm-wave) communication is considered as one of the most important enablers for the fifth generation communication (5G) system to support data rate of Gbps and above. In some scenarios, it is crucial to maintain a line of sight (LOS) link for users enjoying 5G immersive experiences and thus requiring very high data rate. In this paper, we investigate the LOS probability in mm-wave systems. In particular, we study the impact of access point (AP) and blockage height on the LOS probability and propose a solution to effectively enhance the LOS coverage by using high-rise APs on top of low-rise APs normally installed on street furniture, e.g., lamp poles. Two deployment options are explored: 1) irregular deployment and 2) regular deployment, where LOS probability is derived for both cases. Simulation results show that the impact of AP height on LOS probability is significant and using coordinated high-rise APs jointly deployed with low-rise APs will substantially improve the LOS probability.

*Keywords—5G; millimeter wave; line of sight (LOS); non line of sight (NLOS); small cells*

## I. Introduction

With the 5G research well underway in many parts of the world, there is a growing consensus on the need to supplement envisaged sub 6GHz 5G systems with systems operated in the mm-wave spectrum, the mm-wave spectrum (loosely defined as 6-100GHz) provides distinctly different propagation characteristics [1]. The high path loss (per unit aperture size), and high susceptibility to blockage are two main factors affecting the mm-wave systems' coverage, especially outdoors. In this sense, densely deployed mm-wave small cells with multi-node co-ordination seem a feasible solution to both these issues. The use of coordinated BSs to enhance data rate and coverage of the network has been widely studied under the context of 4G LTE/LTE-A network [2]. Various techniques, e.g., joint transmission, coordinated beamforming, and cooperative communication, have been considered [3]-[4]. In 4G wide area coverage with sub-6GHz systems, the coverage degradation in most cases is gradual. Due to the highly directional transmissions and the severe blockage effects in mm-wave, a more 'device centric' approach is needed in multi-node coordination [5]-[7]. In this 'device centric' approach for co-ordinated mm-wave cells, maintaining the LOS connectivity with the active device will be a significant factor.

Although it has been shown in [8] that high data rate via mm-wave communication can be supported by surprisingly rich NLOS links through reflected paths, there still could be a considerable performance degradation compared to LOS links. It becomes more crucial to maintain a LOS link for users requiring very high data rate to support the 5G immersive experiences [9]. In this regard, we focus on LOS coverage in this paper and study the LOS coverage enhancement provided by multi-node coordination. In particular, we consider the coordination between two sets of mm-wave APs: 1) low-rise APs installed on street furniture; and 2) high-rise APs installed on high buildings, and try to answer the following three questions:

1. How many LOS APs can be observed by a typical user equipment (UE) in this specific network scenario?
2. What is the probability that the user is associated with a LOS mm-wave AP (therefore covered by a LOS connection)?
3. What is the improvement in coverage when coordination among two sets of APs is considered?

The rest of the paper is organized at follows. Section II summarizes a simplified LOS probability model. Section III further extends this model by taking AP height into consideration, where both random and regular AP deployment options are explored. The simulation parameters and evaluation results of LOS probability are presented in section IV and the last section concludes the paper.

## II. LOS Coverage Model

We follow the framework proposed in [10], where it has been shown that in a dense mm-wave cellular network where the APs form a homogeneous Poisson point process (PPP), for a typical UE that is assumed to be connected to the AP with the smallest path loss, the probability that a link of length $x$ is LOS is described as a general LOS probability function $p(x)$

$$p(x) = e^{-\beta x}, \qquad (1)$$

where $\beta$ is a parameter determined by size and density of blockages. The probability that the user is associated with a LOS AP is then given as

$$A_L = B_L \int_0^\infty e^{-2\pi\lambda \int_0^{\psi_L(x)} (1-p(t))t\,dt} f_L(x)\,dx, \qquad (2)$$

where λ is the density of PPP APs, $B_L$ is the probability that a user has at least one LOS AP, $f_L(x)$ is the between a user and the nearest LOS base station and we have

$$B_L = 1 - e^{-2\pi\lambda \int_0^x r p(r) dr},$$

$$f_L(x) = 2\pi\lambda x p(x) e^{-2\pi\lambda \int_0^x r p(r) dr} / B_L, \quad (3)$$

$$\psi_L(x) = \left(\frac{C_N}{C_L}\right)^{1/\alpha_N} x^{\alpha_L/\alpha_N}.$$

Here $C_N$ and $C_L$ are the intercepts of the LOS and NLOS path loss formulas, respectively, and $\alpha_L$ and $\alpha_N$ are the path loss exponents for LOS and NLOS, respectively.

The multiple LOS APs available in a small cell range allows us to exploit the coordination among multiple APs to improve the coverage of a typical UE, which is not taken into consideration in the LOS probability eq. (2). In the work that follows, we will be looking at possible coordination schemes for LOS coverage improvement; in particular, we will investigate the impact of AP height on the LOS probability and explore the potential of using two sets of APs with different heights to enhance the LOS coverage.

### III. IMPACT OF HEIGHT ON LOS PROBABILITY

The height of APs has significant impact on the probability of LOS links as shown in Fig. 1, where both building A and B are in between the AP and the UE but the LOS link is blocked by the higher building A.

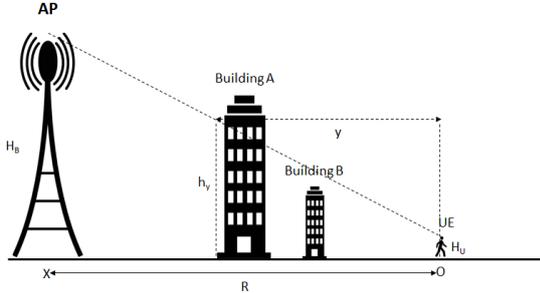

Fig. 1. Blockage by high building

We study two different deployment options: 1) irregular deployment where the AP deployment follows a homogeneous Poisson point process (PPP) and 2) regular deployment where the AP deployment is the conventional hexagon pattern.

*A. Irregular Deployment*

Considering the model in Fig. 1, the building A will block the LOS link as long as its height $h$ is larger than $h_y$. The blocking probability can be expressed as

$$P_{blk} = \int_0^R f(y) P(h > h_y) dy \quad (4)$$

where $f(y)$ is the distribution function of the distance between the building and the UE and $h_y$ can be expressed as

conditional probability density function of the distance

$$h_y = \frac{y H_B + (R-y) H_U}{R},$$

where $H_B$ is the AP height, $H_U$ is the UE height and $R$ is the distance between the AP and the UE.

Assuming the distance between the buildings and the UE follows uniform distribution in the range of [0,R], the blocking probability can be rewritten as

$$\begin{aligned} P_{blk} &= \frac{1}{R} \int_0^{\frac{R(H_{max}-H_U)}{H_B-H_U}} P(h > h_y) dy \\ &= \frac{1}{R} \int_0^{\frac{R(H_{max}-H_U)}{H_B-H_U}} (1 - P(h \leq h_y)) dy \\ &= \frac{1}{R} \int_0^{\frac{R(H_{max}-H_U)}{H_B-H_U}} \left(1 - \int_0^{h_y} f(h) dh\right) dy \\ &= \frac{(H_{max}-H_U)}{H_B-H_U} - \frac{1}{R} \int_0^{\frac{R(H_{max}-H_U)}{H_B-H_U}} \int_0^{h_y} f(h) dh dy \end{aligned} \quad (5)$$

where $f(h)$ is the distribution function of the building height. We assume that the height of buildings follows uniform distribution in the range of [0, $H_{max}$], where the lower bound of the blocking building height is chosen as 0 for simplicity. With this assumption, we have

$$\begin{aligned} P_{blk} &= \frac{(H_{max}-H_U)}{H_B-H_U} - \frac{1}{RH_{max}} \int_0^{\frac{R(H_{max}-H_U)}{H_B-H_U}} \int_0^{h_y} dh dy \\ &= \frac{(H_{max}-H_U)}{H_B-H_U} - \frac{1}{RH_{max}} \int_0^{\frac{R(H_{max}-H_U)}{H_B-H_U}} \frac{y H_B + (R-y) H_U}{R} dy \\ &= \frac{(H_{max}-H_U)}{H_B-H_U} - \frac{H_{max}^2 - H_U^2}{2H_{max}(H_B-H_U)} \\ &= \frac{(H_{max}-H_U)^2}{2H_{max}(H_B-H_U)} \end{aligned} \quad (6)$$

When height is taken into consideration, $\beta$ should be scaled by $P_{blk}$ as $\beta' = \beta P_{blk}$ [11]. Therefore, according to eq. (2), the UE LOS association probability is given

$$\begin{aligned} P_{LOS} &= \int_0^\infty e^{-2\pi\lambda \int_0^{\psi_L(x)} (1-p(t)) t dt} 2\pi\lambda x e^{-\beta P_{blk} x} e^{-2\pi\lambda U(x)} dx \\ &= 2\pi\lambda \int_0^\infty x e^{-2\pi\lambda (Y(\psi_L(x)) + U(x)) - \beta P_{blk} x} dx \end{aligned} \quad (7)$$

where

$$Y(x) = \frac{x^2}{2} - U(x)$$

$$\begin{aligned} U(x) &= \int_0^x r p(r) dr = \int_0^x r e^{-\beta P_{blk} r} dr \\ &= \frac{1}{(\beta P_{blk})^2} - \frac{1}{\beta P_{blk}} e^{-\beta P_{blk} x} \left(\frac{1}{\beta P_{blk}} + x\right) \end{aligned}.$$

*B. Regular Deployment*

The above analysis is based on the assumption that the deployment of APs follows PPP model and the cell radius varies. However, the AP deployment might follow the regular hexagon pattern and therefore the cell radius is fixed. In such a case, we can extend the analysis by assuming the blocking buildings follows uniform distribution from cell centre to cell edge in the range of [0, $R_c$], where $R_c$ is the cell radius. Following the similar derivation as in eq. (4) to (6), the blocking probability can be expressed as

$$P_{blk} = \frac{R(H_{max} - H_U)^2}{2R_c H_{max}(H_B - H_U)}. \quad (8)$$

This equation shows that the blocking probability increases linearly when the user moves away from the AP, which implies that the cell edge users can be more easily blocked.

We then consider the regular hexagon deployment shown in Fig. 2 and extend our analysis to a 2 dimensional space.

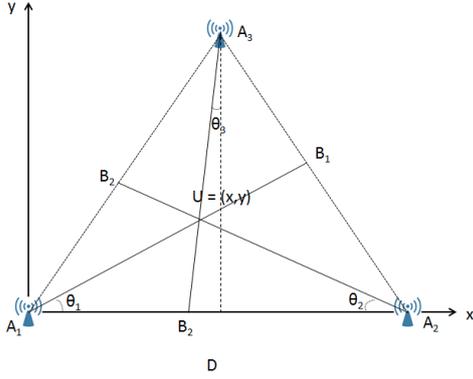

Fig. 2. Regular deployment

As aforementioned, we assume the APs follow hexagon deployment and only consider the users inside the identified triangular area since the users outside this area can be taken care of by other triangular areas. Moreover, only the left side of this triangular area is considered because the blocking or LOS probabilities are symmetrical against the central dotted line. The inter-cell distance is assumed to be $D$ and the cell radius $R_c$ is $D/\sqrt{3}$. The user location is assumed to be at the point $U = \{x, y\}$. The angle between the user and the APs can be expressed as

$$\theta_1 = \arg\tan\left(\frac{y}{x}\right),$$
$$\theta_2 = \arg\tan\left(\frac{y}{D-x}\right), \quad (9)$$
$$\theta_3 = \arg\tan\left(\frac{\frac{D}{2}-x}{\frac{\sqrt{3}}{2}D-y}\right),$$

respectively. The distance between the user and the APs is given as

$$R_1 = |A_1 U| = \sqrt{x^2 + y^2},$$
$$R_2 = |A_2 U| = \sqrt{(D-x)^2 + y^2}, \quad (10)$$
$$R_3 = |A_3 U| = \sqrt{\left(\frac{D}{2} - x\right)^2 + \left(\frac{\sqrt{3}}{2}D - y\right)^2},$$

respectively. Based on eq. (8), the blocking probability of each AP is given as

$$P_{blk,1} = \frac{R_1(H_{max} - H_U)^2}{2|A_1 B_1|H_{max}(H_B - H_U)},$$
$$P_{blk,2} = \frac{R_2(H_{max} - H_U)^2}{2|A_2 B_2|H_{max}(H_B - H_U)}, \quad (11)$$
$$P_{blk,3} = \frac{R_3(H_{max} - H_U)^2}{2|A_3 B_3|H_{max}(H_B - H_U)},$$

respectively, where

$$|A_1 B_1| = \frac{\sqrt{3}D}{2\sin\left(\frac{2\pi}{3} - \theta_1\right)},$$
$$|A_2 B_2| = \frac{\sqrt{3}D}{2\sin\left(\frac{2\pi}{3} - \theta_2\right)},$$
$$|A_3 B_3| = \frac{\sqrt{3}D}{2\sin\left(\frac{\pi}{2} - \theta_1\right)}.$$

As aforementioned, the LOS probability is $e^{-\beta' R}$ where $\beta' = \beta P_{blk}$. The probability that at least one LOS link exists can be expressed as

$$P_{LOS} = 1 - \left(1 - e^{-\beta P_{blk,1} R_1}\right)\left(1 - e^{-\beta P_{blk,2} R_2}\right)\left(1 - e^{-\beta P_{blk,3} R_3}\right)$$
(12)

### C. $P_{LOS}$ with Coordination between Two Sets of APs

In reality, large amount of low-rise APs are normally installed on street furniture. However LOS links could be easily blocked especially for the cell edge users. The blockage of LOS links will significantly affect the users enjoying immersive multi-media experiences such as UHD and virtual reality gaming. We propose a solution to such a situation. That is to install a small amount of high-rise APs on high building to maintain the LOS connection when it cannot be provided by low-rise APs, thus enhancing the 5G immersive experiences by coordination between these two sets of APs.

When both high- and low-rise APs are considered, the LOS probability can be expressed as

$$P_{LOS} = 1 - (1 - P_{LOS,l})(1 - P_{LOS,h}), \quad (13)$$

where $P_{LOS,l}$ and $P_{LOS,h}$ are LOS probability for low- and high-rise APs, respectively.

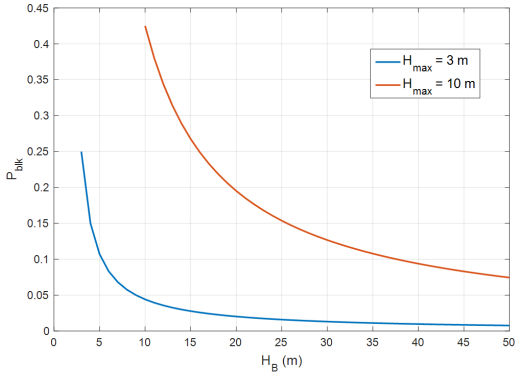

Fig. 3. Blocking probability

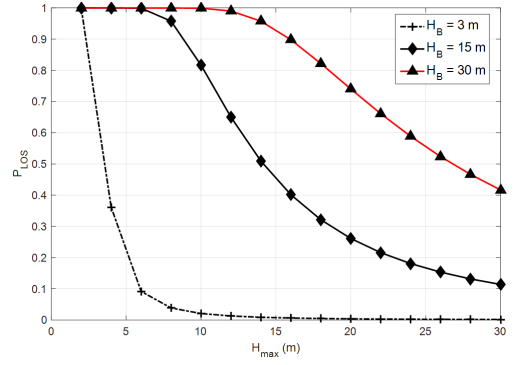

Fig. 5. LOS probability with increased blocking building height (average cell radius = 100 m)

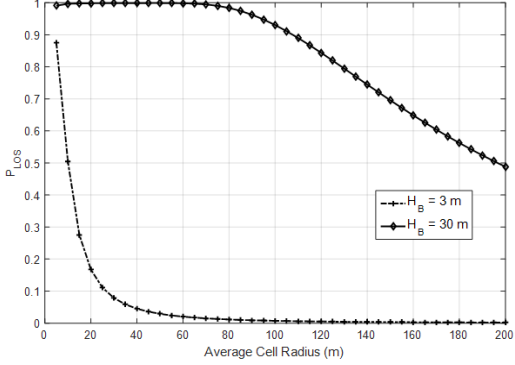

Fig. 4. LOS association probability for $H_B$ = 3 and 30 m, respectively ($H_{max}$ = 15 m)

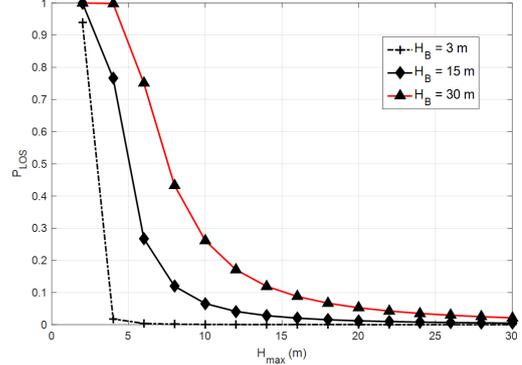

Fig. 6. LOS probability with increased blocking building height (average cell radius = 500 m)

## IV. NUMERICAL RESULTS

In this section, we show some numerical results to demonstrate the impact of height of AP on LOS association probability and the potential improvement of LOS coverage using the proposed multi-node coordination. The system parameters are given in Table-I and we assume $C_N = C_L$ [12].

Table-I System Parameters

| Simulation Parameter | Value |
| --- | --- |
| Carrier Frequency | 28 GHz |
| Cell radius (m) | up to 1000 |
| $\alpha_N$ | 2 |
| $\alpha_L$ | 4 |
| $H_U$ (m) | 1.5 |
| $H_{max}$ (m) | 3, 10, 15 and 30 |
| $\beta$ | 0.0709 |

### A. Irregular Deployment

The impact of AP height is clearly shown in Fig. 3. With maximum blocking building height 3 m, the LOS blocking probability can be reduced from 0.25 to below 0.05 if the AP height is increased from 3 m to 10 m. Even with higher maximum blocking building height 10 m, the blocking probability can be reduced from over 0.4 to below 0.1 if the AP height is over 40 m.

Fig. 4 shows the LOS association probability for different AP heights with increased average cell radius. It can be easily seen that the height of the AP and cell density play crucial roles in determining the UE LOS association probability. When a lower AP is installed on the street furniture, e.g., lamp post, with height 3 m, the LOS association probability reduces rapidly with increased cell radius, i.e., reduced cell density. However, when an AP is installed in a high building with height 30 m, the LOS association probability decreases much slower. It means that the high-rise APs can always provide significantly high LOS coverage even with low cell density.

It is also interesting to see the impact of the blocking building height on the LOS probability as shown in Fig. 5 and Fig. 6.

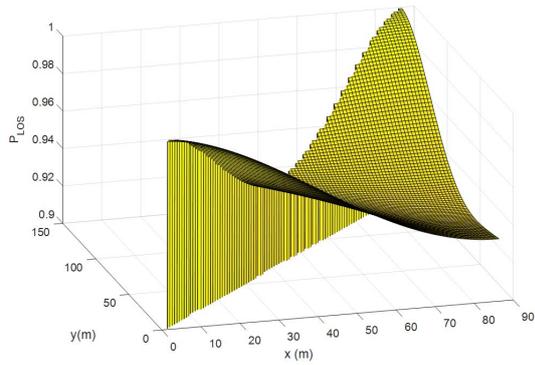

Fig. 7. LOS probability $P_{LOS}$ ($R_c$ = 100m, $H_B$ = 3 m)

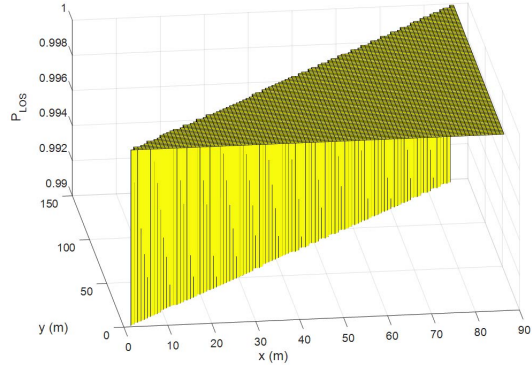

Fig. 8. LOS probability $P_{LOS}$ ($R_c$ = 100m, $H_B$ = 30 m)

As can be seen, the LOS probability is sensitive with respect to the blocking building height and low-rise AP is more likely to be affected than high-rise AP. For example, for average cell radius 100 m, when the blocking building height is increased from 2 m to 6 m, the LOS probability of low-rise AP ($H_B$=3 m) reduces from almost 1 to 0.1. In contrast, for high-rise AP ($H_B$ =30 m) the LOS keeps unchanged. Moreover, the LOS probability gets more sensitive with respect to blocking building height when the average cell radius increases. With larger average cell radius, a small increase in blocking building height will cause significant decrease in LOS probability. This is reasonable because when the user moves away from the AP they are more likely to be blocked.

### B. Regular Deployment

With the same evaluation parameters, the LOS probabilities of regular deployment are shown in Fig. 7 and Fig. 8. Here we only show the left part of the triangular areas shown in Fig. 2 and therefore two APs are located at {0, 0} and {D/2, D/$\sqrt{3}$ }.

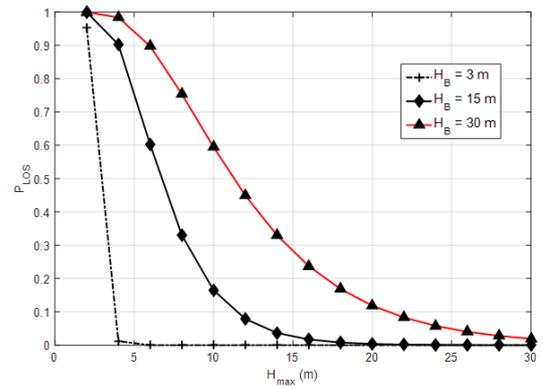

Fig. 9. LOS probability with increased blocking building height (cell radius = 500 m)

There are following important observations from these figures:

- The users are more likely to have LOS links when they are close to the APs and the LOS probability decreases when the users move away from the APs;
- Increasing the AP height from 3 to 30 m significantly increases the LOS probability. For AP height 30 m, the LOS probability keeps at a very high level (above 0.99).

These two observations comply with previous observations for the irregular deployment scenario.

Fig. 9 shows the LOS probability with increased maximum height of blocking buildings. The same trends as irregular deployment scenario are observed. It should be noted that the LOS probability shown here is the worst case LOS probability, i.e., the lowest point in Fig. 7 and Fig. 8 when the user is far away from the APs. It is possible that the users close to the AP could still have large LOS probability even with low height APs and large radius.

### C. Joint Deployment of Low- and High-Rise APs

From the simulation results, we can see that the LOS probability can be significantly enhanced by increasing the height of APs. However, high-rise APs cannot be deployed as dense as low-rise APs because high buildings are not always as available as street furniture such as lamp poles. Therefore we consider installing only small number of high-rise APs on top of large amount of low-rise APs and show the LOS coverage improvement due to the joint deployment.

In irregular deployment, we assume that the height of the low- and high-rise APs is 3 m and 30 m, respectively, and blocking building height is 3 m. We also assume that the low-rise AP cell radius is 100 m and the LOS probability with low-rise APs only is 0.93. Then we assume one high-rise AP is installed for each 100 low-rise APs. The overall LOS probability is improved from 0.93 to almost 1 based on eq. (13). If the blocking building height is increased from 3 m to 10 m and only low-rise APs are deployed, the LOS probability is only 0.018, which means LOS coverage is almost not available. However, if high-rise APs with radius 300 m are deployed jointly, the overall LOS coverage is improved from 0.018 to 0.833. In such a case, only one high-rise AP needs to be installed for each 9 low-rise APs. For regular deployment with the same assumptions, the LOS probability of low-rise APs only is 0.8564 but if one high-rise AP is installed for each 100 low-rise APs, the LOS probability can be enhanced to 0.9998 because of this joint deployment.

In Table-II, we show that for a target overall LOS probability 0.95, how many high-rise APs (30 m) need to be installed together with each 100 low-rise APs (3 m) in regular deployment scenario. It can be seen that only very few high-rise APs are needed to enhance the LOS coverage when the blocking building height is low. However, with increased blocking building height, more and more high-rise APs are needed to provide target LOS coverage.

Table-II Number of high-rise APs

| Blocking Building Height (m) | $P_{LOS}$ with Low-Rise AP only | Number of High-rise APs | $P_{LOS}$ with Joint Deployment |
|---|---|---|---|
| 3 | 0.93 | 1 | 0.9990 |
| 5 | 0.44 | 3 | 0.9578 |
| 10 | 0.02 | 25 | 0.9636 |
| 15 | 0.0005 | 100 | 0.9513 |

V. CONCLUSIONS AND FUTURE WORKS

In this paper, we investigate the LOS probability for mm-wave communication systems considering both irregular and regular deployed mm-wave APs. The LOS probability is derived by taking AP and blockage height into account. It has been shown that the height of APs could significantly impact the LOS probability. We propose a solution using high-rise APs to coordinate with low-rise APs and the simulation results show that the joint deployment and coordination could significantly improve the LOS coverage and it is possible to use only one or a few jointly deployed high-rise APs to achieve this improvement when the blockage is not very high.

In further developing this research, optimum combination ratios for low-rise and high-rise APs will be studied, for different urban scenarios. Also we will bring the interference effects to the analysis, as interference would play a significant role in the overall deployment options. Finally we will investigate the best co-ordination strategies between low rise and high rise APs, with the twin aims of maximizing coverage/capacity and minimizing interference.

ACKNOWLEDGEMENTS

The European Commission funding under H2020-ICT-14-2014 (Advanced 5G Network Infrastructure for the Future Internet, 5G PPP), and project partners: Samsung, Ericsson, Aalto University, Alcatel-Lucent, CEA LETI, Fraunhofer HHI, Huawei, Intel, IMDEA Networks, Nokia, Orange, Telefonica, Bristol University, Qamcom, Chalmers University of Technology, Keysight Technologies, Rohde & Schwarz, TU Dresden are acknowledged.